# Inverse Uncertainty Quantification using the Modular Bayesian Approach based on Gaussian Process, Part 2: Application to TRACE

Xu Wu[a,*], Tomasz Kozlowski[a], Hadi Meidani[b] and Koroush Shirvan[c]

[a]Department of Nuclear, Plasma and Radiological Engineering, University of Illinois at Urbana-Champaign, Urbana, IL, USA

[b]Department of Civil and Environmental Engineering, University of Illinois at Urbana-Champaign, Urbana, IL, USA

[c]Department of Nuclear Science and Engineering, Massachusetts Institute of Technology, Cambridge, MA, USA

*Phone: (+1) 217-979-7432, Email: xuwu2@illinois.edu

**Abstract**

Inverse Uncertainty Quantification (UQ) is a process to quantify the uncertainties in random input parameters while achieving consistency between code simulations and physical observations. In this paper, we performed inverse UQ using an improved modular Bayesian approach based on Gaussian Process (GP) for TRACE physical model parameters using the BWR Full-size Fine-Mesh Bundle Tests (BFBT) benchmark steady-state void fraction data. The model discrepancy is described with a GP emulator. Numerical tests have demonstrated that such treatment of model discrepancy can avoid over-fitting. Furthermore, we constructed a fast-running and accurate GP emulator to replace TRACE full model during Markov Chain Monte Carlo (MCMC) sampling. The computational cost was demonstrated to be reduced by several orders of magnitude.

A sequential approach was also developed for efficient test source allocation (TSA) for inverse UQ and validation. This sequential TSA methodology first selects experimental tests for validation that has a full coverage of the test domain to avoid extrapolation of model discrepancy term when evaluated at input setting of tests for inverse UQ. Then it selects tests that tend to reside in the unfilled zones of the test domain for inverse UQ, so that one can extract the most information for posterior probability distributions of calibration parameters using only a relatively small number of tests. This research addresses the "lack of input uncertainty information" issue for TRACE physical input parameters, which was usually ignored or described using expert opinion or user self-assessment in previous work. The resulting posterior probability distributions of TRACE parameters can be used in future uncertainty, sensitivity and validation studies of TRACE code for nuclear reactor system design and safety analysis.

## 1. Introduction

The significance of Uncertainty Quantification (UQ) has been widely recognized in the nuclear community and numerous publications have been devoted to UQ methods and applications in response to the Best Estimate Plus Uncertainty (BEPU) methodology [1][2][3][4]. However, the concept of UQ in the nuclear community generally means "forward UQ", which is the process to propagate input uncertainties to the Quantity-of-Interests (QoIs) via the computer codes. Forward UQ requires the input uncertainty information, such as the statistical moments, Probability Density Functions (PDFs), upper and lower bounds, etc. "Expert opinion" or "user self-assessment" have been widely used to specify such information in previous uncertainty and sensitivity studies. Such ad-hoc specifications are unscientific and lack mathematical rigor, even if they have been considered reasonable for a long time.

Inverse UQ can be used to tackle the "lack of input uncertainty information" issue, which is a process to quantify the input uncertainties given experimental data. In our companion paper [5], we discussed the connection and difference between inverse UQ and calibration. In brief, deterministic calibration only results in point estimates of best-fit input parameters, while Bayesian calibration and inverse UQ target at quantifying the uncertainties in these input uncertainties. Since measurement data are usually insufficient to inform us about the "true" or "exact" values of the calibration parameters, uncertainties in calibration parameters should be quantified to prevent over-confidence in the calibration process. Besides the subtle differences between Bayesian calibration and inverse UQ discussed in [5], they can be treated as the same concept in most cases.

One of the earlier works on inverse UQ[1] in nuclear engineering was the "Circé method" presented in [6], which was developed to quantify the uncertainties in the closure laws of Cathare 2 code. The Circé method implemented the Expectation-Maximization (EM) algorithm in conjunction with the Maximum Likelihood Estimate (MLE) method. This approach was later extended to a richer mathematical framework which included Maximum a Posteriori (MAP) [7]. In [8], MLE, MAP and Markov Chain Monte Carlo (MCMC) algorithms were used to quantify the uncertainty of two physical models used in TRACE: subcooled boiling heat transfer coefficient (HTC) and interfacial drag coefficient. However, the applications of these methods are limited by several strong assumptions: (1) the relation between the QoIs and the input parameters were assumed to be linear; (2) the input parameters were assumed to follow normal distributions; (3) local sensitivity analysis was required to provide the necessary inputs for MLE.

Cacuci and Arslan [9] applied a predictive modeling procedure to reduce the uncertainties in calibration parameters and time-dependent boundary conditions in the large-scale reactor analysis code FLICA4 based on the BFBT benchmark, yielding best-estimate predictions of axial void fraction distributions. Bui and co-workers [10][11] proposed the concept of "total model-data integration" which is based on the theory of Bayesian calibration and a mechanism to accommodate multiple data-streams and models. Such concept allows assimilation of heterogeneous multivariate data in comprehensive calibration and validation of computer models. This approach was demonstrated on the cases of subcooled boiling two-phase flow in nuclear thermal-hydraulics (TH) analysis. Bachoc *et al.* [12] applied the Bayesian calibration approach to the TH code FLICA 4 in a single-phase friction model. The model discrepancy was modeled with a Gaussian Process (GP). This work made inference of the model error for each new potential experimental point, extrapolated from what had been learnt from the available experimental data. The computer code predictive capability was reported to be improved based on the tested case. In another work [13], Bayesian calibration was applied to calibrate the reflood model parameters in TRACE. Furthermore, this work considered multivariate time-dependent outputs and the model discrepancy. However, no results for the quantified model discrepancy were presented and its extrapolation to the validation and prediction domain was not discussed.

GP emulator-based Bayesian calibration was also applied to fuel performance codes. Higdon *et al.* [14] used the full Bayesian approach to inversely quantify the uncertainties in four tuning parameters of the FRAPCON code based on fission gas release data from 42 experiments. The measurement uncertainty and the model discrepancy term were quantified simultaneously. Kriging-based calibration was used to quantify the calibration parameters in the fission gas behavior model of fuel performance code BISON [15][16]. Single experiment was used in [15] while multiple integral experiments were used in [16]. However, they only resulted in an optimized set of fission gas behavior model parameters rather than posterior distributions. Another example of kriging-based inverse UQ can be found in [17], where the authors proposed a method for cases when time-series data is used for inverse UQ. Principal Component Analysis (PCA) was used to project the original time-series data to the principal component subspace. Inverse UQ was then performed on the subspace to avoid convergence issues using the original measurement data.

Surrogate-based calibration was also used in nuclear engineering applications other than TH and fuel performance. Stripling *et al*. [18] developed a method for calibration and data assimilation using the Bayesian Multivariate Adaptive Regression Splines (MARS) emulator as a surrogate for the computer code. Their method started with sampling of the uncertain input space. The emulator was then used to assign weights to the samples which were applied to produce the posterior distributions of the inputs. This approach was applied to the calibration of a Hyades 2D model of laser energy deposition in beryllium. The major difference of this approach with MCMC-based Bayesian calibration is that, it generated samples beforehand and the candidate acceptance routine in MCMC sampling was replaced with a weighting scheme. Note that such approach did not include a model discrepancy term. Yurko *et al*. [19] used the Function Factorization with Gaussian Process (FFGP) priors model to emulate the behavior of computer codes. Calibration of a simple friction-factor example using a Method of Manufactured Solution (MMS) with synthetic observational data was used to demonstrate the key properties of this method. This approach is better suited for the emulation of complex time series outputs. Wu and Kozlowski [20] used generalized Polynomial Chaos Expansion (PCE) to construct surrogate models for inverse UQ of a point kinetics coupled with lumped parameter TH feedback model, also with synthetic measurement data. The developed approach was demonstrated to be capable of identifying the (pre-specified) "true" values of calibration parameters and greatly reducing the computational cost.

---

[1] Most of the mentioned previous work did not use the term "inverse UQ". But according to our definition they are equivalent to inverse UQ.

In a companion paper [5], we presented the detailed theory and procedure to perform inverse UQ. We also proposed an improved modular Bayesian approach based on GP. In this paper, the proposed method is implemented to system TH code TRACE [21] using experimental data from the OECD/NEA BWR Full-size Fine-Mesh Bundle Tests (BFBT) benchmark [22]. Note that in a previous paper [23], we also performed inverse UQ of the same code with the same benchmark but using surrogate model constructed by Sparse Grid Stochastic Collocation (SGSC). SGSC surrogate model can also greatly reduce the computational cost. However, unlike GP, it cannot be used to represent the model discrepancy during inverse UQ. In this work, we have greatly improved the application in the following aspects:

1) GP is used to construct metamodel for TRACE code, which requires even less TRACE runs than SGSC used in [23]. Furthermore, as shown in our companion paper [5], GP metamodel provides Mean Square Error (MSE) of its prediction which is essentially the "code uncertainty" (see Section 2.3 in [5]).

2) The previous work [23] only used 8 experiment tests from BFBT benchmark test assembly No.4. In this work we will use all the 86 test cases.

3) Only void fraction data from upper elevations were used for inverse UQ in [23] because void fraction measurements at lower elevations are sometimes physically wrong (negative). In this work, we will use all the void fraction measurements to show our "respect" to the reported data, considering that those negative void fractions are very close to zero.

4) The previous work [23] did not consider model discrepancy during inverse UQ. Therefore, the results are likely to be over-fitted to the selected test cases. In this work, we will describe the model discrepancy term with GP to avoid over-fitting, following the steps outlined in [5].

An unresolved issue for inverse UQ is "test[2] source allocation (TSA)". TSA is the process to divide a set of given experimental data into training (calibration, or inverse UQ) and testing (validation) sets, as the same data should not be used for both purposes. Very little previous research on Bayesian calibration dealt with TSA. Some researchers simply used random selection [24]. In our previous work on inverse UQ of TRACE [23], we separated tests by certain ad-hoc criteria, for example, tests with high pressure and high power were selected for inverse UQ.

A data partitioning methodology adapted from cross-validation was presented in [25]. The method aimed at separating legacy data for calibration and validation purposes. It considered all possible partitions and tried to find the optimal partition satisfying the following desiderata: (1) the model is sufficiently informed by calibration tests, (2) validation tests challenge the model as much as possible with respect to the QoIs. It should be noted that this method is extremely expensive. For an original data set of size $N$, the number of inverse (calibration) problems to solve is $(2^N - 2)$. For example, for 10 experimental tests, 1022 inverse problems need to be solved. This approach is not practical in our application which has 86 experimental test cases. Recently, a test selection methodology was developed [26] that involves an optimization framework for integrating calibration and validation data to make a prediction. In this approach, the TSA is motivated by uncertainty reduction in prediction. However, this method is designed for a situation where the actual experiments have not been conducted yet, while in the present case we are interested in separating data based on experiments that have already been performed.

In this paper, we proposed a sequential approach for TSA. This algorithm includes three steps: (1) selecting an initial set of validation data from all the tests; (2) selecting an initial set of inverse UQ data after removing the initial validation set; (3) sequentially adding test cases for inverse UQ from the remaining tests. This algorithm guarantees that the tests used for validation have a maximum coverage of the test domain. Meanwhile, the tests used for inverse UQ have the lowest discrepancy which means that they explore the test space to the largest extent.

This paper is organized in the following way. Section 2 briefly introduces the necessity to perform inverse UQ for TH code closure models, the TRACE code and the BFBT benchmark. The sequential approach for TSA is described in Section 3. In Section 4, we present the details about how the TRACE model discrepancy is modeled. Section 5 demonstrates the process to construct and validate the metamodel for TRACE which will be used later for MCMC sampling during inverse UQ. The results for inverse UQ are included in Section 6. Section 7 concludes the paper.

---

[2] Unless explicitly specified, in this paper "test" generally means "experimental test", not the one used in "train vs. test" in computer modelling.

## 2. Problem Description

For nuclear reactor best-estimate system TH codes, significant uncertainties come from the closure laws which are used to describe the transfer terms in the balance equations. These physical models govern the mass, momentum and energy exchange between the fluid phases and surrounding medium, varying according to the type of a two-phase flow regime. When the closure models were originally developed, their accuracy and reliability were studied with a particular experiment. However, once they are implemented in a TH code as empirical correlations and used for prediction of different physical systems, the accuracy and uncertainty characteristics of these correlations are no longer known to the user. In the current research, we focus on the physical model parameters related to these models. Previously in the uncertainty and sensitivity study of such codes, physical model parameter uncertainty distributions are simply ignored, or described using expert opinion or user self-evaluation. This necessitates a framework that accurately quantifies the uncertainties associated with physical model parameters of best-estimate system TH codes.

### 2.1. Problem overview

TRACE version 5.0 Patch 4 [21] includes options for user access to 36 physical model parameters. The details of these parameters can be found in Appendix A. For forward uncertainty propagation, the users are free to perturb these parameters by addition or multiplication according to their personal opinion or expert judgment. The work presented in this paper will inversely quantify the uncertainties of these parameters based on experimental data. All quantified uncertainties will be multiplicative factors of the nominal values.

The international OECD/NRC BFBT benchmark, based on the Nuclear Power Engineering Corporation (NUPEC) database [22], was created to encourage advancement in sub-channel analysis of two-phase flow in rod bundles, which has great relevance to the nuclear reactor safety evaluation. In the frame of the BFBT test program, single- and two-phase pressure losses, void fraction, and critical power tests were performed for steady-state and transient conditions. Detailed description of BFBT benchmark can be found in [22][23]. In the present work, steady-state void fraction data from BFBT test assembly 4 is used, which consists of 86 tests. Cross-sectional averaged void fractions were measured at four different axial locations, hereafter referred to as VoidF1, VoidF2, VoidF3 and VoidF4 respectively from lower to upper positions.

The selected assembly 4 void fraction data then went through data correction and selection processes, which are described in Appendix B. Eventually 78 tests (78*4 = 312 void fraction observations) will be used in the following study. Furthermore, 36 uncertain physical model parameters make it very difficult to perform inverse UQ for TRACE. The number of training samples generally increases exponentially with the dimension, creating a challenge commonly referred to as the "curse of dimensionality" in literature [23]. Therefore, we need to perform dimensional reduction for the current problem before performing inverse UQ. Previous study [23] combined local and global sensitivity study to identify the significant parameters for this problem. Table 1 shows the five parameters selected after dimensional reduction. The nominal values for all these calibration parameters are 1.0 since they are multiplication factors (i.e., all the parameters are normalized with respect to their nominal values). The prior ranges are chosen as [0, 5] for all the parameters which will be used later in design of computer experiments to build the GP metamodels. The prior ranges are chosen to be wide to reflect the ignorance of these parameters. Posterior ranges resulting from inverse UQ are expected to be much narrower than prior ranges, indicating that the knowledge in these parameters has been improved given physical observations.

Table 1. Selected TRACE physical model parameters after sensitivity analysis

| Parameter (multiplication factors) | Representation | Uniform range | Nominal |
| --- | --- | --- | --- |
| Single phase liquid to wall HTC | P1008 | (0.0, 5.0) | 1.0 |
| Subcooled boiling HTC | P1012 | (0.0, 5.0) | 1.0 |
| Wall drag coefficient | P1022 | (0.0, 5.0) | 1.0 |
| Interfacial drag (bubbly/slug Rod Bundle - Bestion) coefficient | P1028 | (0.0, 5.0) | 1.0 |
| Interfacial drag (bubbly/slug Vessel) coefficient | P1029 | (0.0, 5.0) | 1.0 |

### 2.2. Workflow for the investigated problem

Following the notations in the companion paper [5], in this problem: (1) $y^M$ and $y^E$ represent the simulated and measured void fractions (VoidF1, VoidF2, VoidF3 and VoidF4); (2) $y^M(\mathbf{x}, \boldsymbol{\theta})$ is the TRACE code; (3) the set of design variables includes **x** are pressure, mass flow rate, power and inlet temperature; (4) calibration parameters **θ** are P1008, P1012, P1022, P1028 and P1029. Figure 1 shows the improved modular Bayesian approach we proposed in [5]. This flowchart is put here to improve the clarity of the workflow in the application. All the five steps in Figure 1 were explained in the companion paper [5], and Sections 3 - 6 will go through them one-by-one (step 5 validation will be presented in a future work). We will also provide explanations of the major steps in Sections 3 - 6 so that this manuscript is self-contained and understandable without referring too much to the companion theory paper [5].

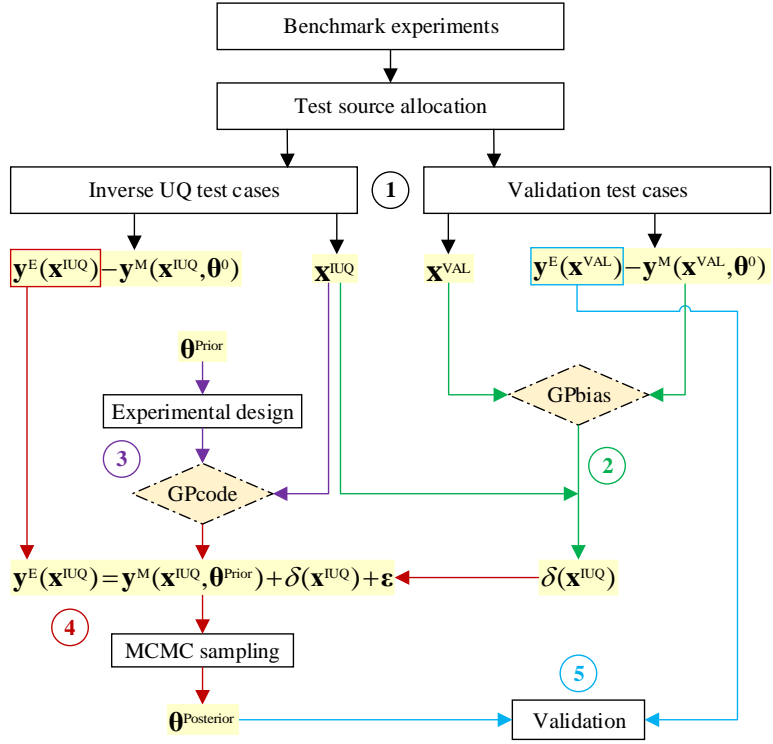

Figure 1: Flowchart of the improved modular Bayesian approach.

## 3. Test Source Allocation

In this section, a sequential approach for efficient TSA is developed (as shown in blocks connected by black arrows in Figure 1). This section starts with a brief introduction of discrepancy measure, which is later used as a measure of the degree of uniformity for the distribution of inverse UQ tests in the whole test domain. The following sub-sections describe the sequential approach for TSA, as well as two algorithms to select initial sets for validation and inverse UQ.

### 3.1. Discrepancy measure

In the companion paper [5], we briefly mentioned low-discrepancy sequences for design of computer experiments. Low discrepancy sequences are deterministic designs constructed to uniformly fill the space. Various discrepancy measures can be used to judge the uniformity quality of the design, see discussions in [27]. Discrepancy measures based on $L_2$ norms can be analytically expressed which makes them the most popular in practice among others. For example, given a design $\mathbf{X}(n) = \{x_k^{(i)}, i = 1,2, \dots, n, k = 1,2, \dots, d\}$ where $n$ is the number of design points and $d$ is the dimension of each design point, the *centered $L^2$ discrepancy* is calculated as:

$$D^2[\mathbf{X}(n)] = \left(\frac{13}{12}\right)^d - \frac{2}{n}\sum_{i=1}^{n}\prod_{k=1}^{d}\left(1 + \frac{1}{2}\left|u_k^{(i)} - \frac{1}{2}\right| - \frac{1}{2}\left|u_k^{(i)} - \frac{1}{2}\right|^2\right)$$
$$+ \frac{1}{n^2}\sum_{i=1}^{n}\sum_{j=1}^{n}\prod_{k=1}^{d}\left(1 + \frac{1}{2}\left|u_k^{(i)} - \frac{1}{2}\right| + \frac{1}{2}\left|u_k^{(j)} - \frac{1}{2}\right| - \frac{1}{2}\left|u_k^{(i)} - u_k^{(j)}\right|\right) \quad (1)$$

Where $\{u_k^{(i)}, i = 1,2,...,n, k = 1,2,...,d\}$ are the normalized values of $\mathbf{X}(n)$ in the interval $[0,1]$. A design sequence with a smaller centered $L^2$ discrepancy has a better coverage of the domain. Another recommended discrepancy measure is the *wrap-around $L^2$ discrepancy* defined in Equation (2), which allows to suppress bound effects [27] (by wrapping the unit cube for each dimension).

$$W^2[\mathbf{X}(n)] = \left(\frac{4}{3}\right)^d + \frac{1}{n^2}\sum_{i=1}^{n}\sum_{j=1}^{n}\prod_{k=1}^{d}\left(\frac{3}{2} - |u_k^{(i)} - u_k^{(j)}|(1 - |u_k^{(i)} - u_k^{(j)}|)\right) \qquad (2)$$

Besides their application in design of computer experiments, researchers have found other applications of low discrepancy sequences. For example, a "sequential validation design" was developed in [27] to select test points for the validation of metamodels. Suppose the metamodel was originally trained based on the sample set $\mathbf{X}_s$. Test points from a low discrepancy sequence $\mathbf{X}_f$ (e.g. Sobol, Halton, Hammersley, etc.) were selected one-by-one according to the criterion that among all the remaining points in $\mathbf{X}_f$, the selected point results in the minimal centered $L^2$ discrepancy after being added to $\mathbf{X}_s$. This design algorithm can avoid the possibility of too strong proximity between training sites and test sites, because it is capable of putting points in the unfilled zones of the training design.

### 3.2. A sequential approach for test source allocation

In the current work, we employ an idea similar to the "sequential validation design" [27] to separate experimental test cases for inverse UQ and validation. Given $N_{\text{test}}$ experimental tests on the test domain $\mathbf{x}^{\text{test}}$, we would like to select $N_{\text{IUQ}}$ tests for inverse UQ and the remaining $N_{\text{VAL}}$ tests to validate the updated model after inverse UQ. Let us denote the inverse UQ domain and validation domain as $\mathbf{x}^{\text{IUQ}}$ and $\mathbf{x}^{\text{VAL}}$, respectively. Then we have:

$$\mathbf{x}^{\text{test}} = \mathbf{x}^{\text{IUQ}} \cup \mathbf{x}^{\text{VAL}}, \quad N_{\text{test}} = N_{\text{IUQ}} + N_{\text{VAL}} \qquad (3)$$

Following the notations in [5], each of the input settings is a $r$-dimensional vector $\mathbf{x} = [x_1, x_2, ..., x_r]^T$ representing $r$ different design variables. In the current work $r = 4$. Figure 2 shows the workflow of the sequential approach for TSA. It includes the following key steps:

- Step 1: find initial set for validation:

    The initial set for validation is denoted as $\mathbf{x}^{\text{VAL,init}}$. The set $\mathbf{x}^{\text{VAL,init}}$ tests are selected from all the tests in $\mathbf{x}^{\text{test}}$. The selection criterion is that the validation experiments should have a full coverage of the test domain. The motivation and solution are explained in Section 3.3.

- Step 2: find initial set for inverse UQ:

    After removing $\mathbf{x}^{\text{VAL,init}}$ from $\mathbf{x}^{\text{test}}$, the remaining tests are defined as $\mathbf{x}^{\text{rest}} = \mathbf{x}^{\text{test}} \setminus \mathbf{x}^{\text{VAL,init}}$. In this step we select initial set for inverse UQ $\mathbf{x}^{\text{IUQ,init}}$ from $\mathbf{x}^{\text{rest}}$. The selection criterion is that the tests in $\mathbf{x}^{\text{IUQ,init}}$ tend to be "far away" from other tests in the domain of interest. The motivation and solution are explained in Section 3.4.

- Step 3: sequentially add more tests for inverse UQ:

    Again we remove $\mathbf{x}^{\text{IUQ,init}}$ from $\mathbf{x}^{\text{rest}}$, that is $\mathbf{x}^{\text{rest}} = \mathbf{x}^{\text{rest}} \setminus \mathbf{x}^{\text{IUQ,init}}$. This step loops through all the tests in $\mathbf{x}^{\text{rest}}$ to add one test each time to $\mathbf{x}^{\text{IUQ}}$. At the beginning, $\mathbf{x}^{\text{IUQ}} = \mathbf{x}^{\text{IUQ,init}}$. Then the algorithm find the test $\mathbf{x}^{(i^*)}$ from $\mathbf{x}^{\text{rest}}$ such that after being added to $\mathbf{x}^{\text{IUQ}}$, the minimum centered $L^2$ discrepancy $D^2[\mathbf{x}^{\text{IUQ}} \cup \mathbf{x}^{(i^*)}]$ or wrap-around $L^2$ discrepancy $W^2[\mathbf{x}^{\text{IUQ}} \cup \mathbf{x}^{(i^*)}]$ is achieved. Next the test $\mathbf{x}^{(i^*)}$ will be moved to the inverse UQ set, i.e. $\mathbf{x}^{\text{IUQ}} = \mathbf{x}^{\text{IUQ}} \cup \mathbf{x}^{(i^*)}, \mathbf{x}^{\text{rest}} = \mathbf{x}^{\text{rest}} \setminus \mathbf{x}^{(i^*)}$.

- Step 4: decision about terminating the sequential approach:

    This step decides whether the desirable number of tests for inverse UQ has been reached $N_{\text{IUQ}} = \lfloor N_{\text{test}} \cdot \alpha \rfloor$. The round-down symbol $\lfloor N \rfloor$ means we take the largest integer that does not exceed $N$. In the current study we choose $\alpha = 0.25$ which means we use about 25% of all the tests for inverse UQ, and all the rest for validation $\mathbf{x}^{\text{VAL}} = \mathbf{x}^{\text{test}} \setminus \mathbf{x}^{\text{IUQ}}$. The reason for such choice is explained in Section 3.3. If the

desirable number is reached, we proceed to perform inverse UQ with $\mathbf{x}^{\text{IUQ}}$. Otherwise, repeat step 3 to add another test for $\mathbf{x}^{\text{IUQ}}$.

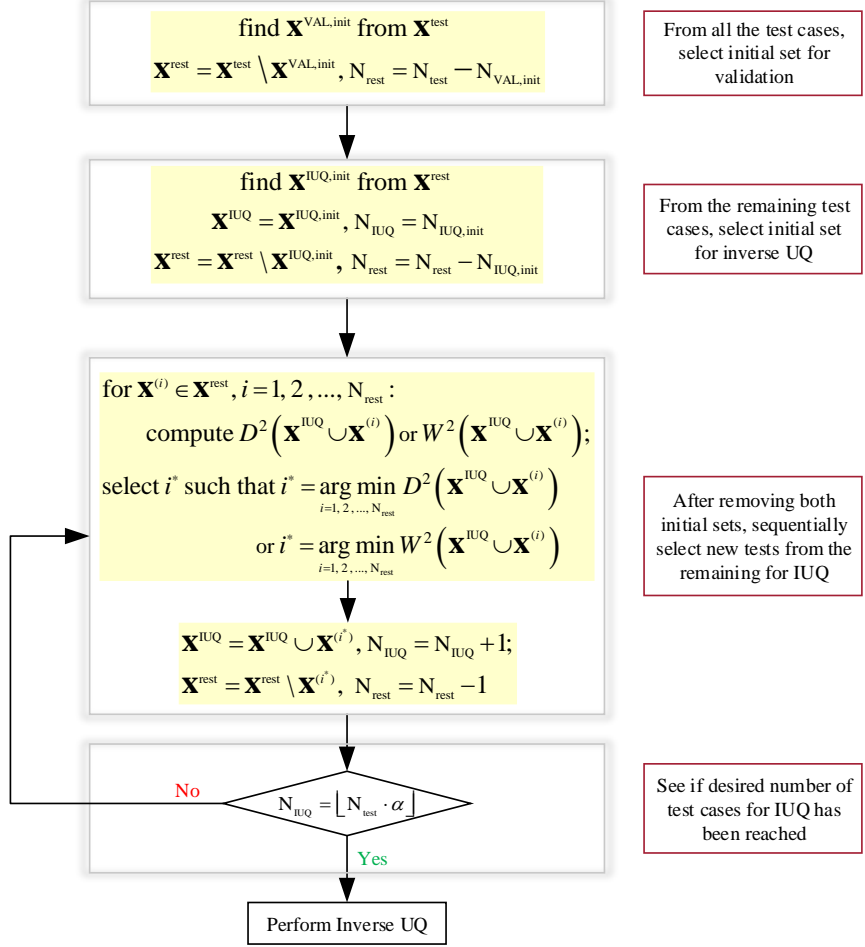

Figure 2: The sequential approach for test source allocation.

By following the steps outlined in Figure 2, $\mathbf{x}^{\text{IUQ,init}}$ and $\mathbf{x}^{\text{VAL,init}}$ will be subsets of $\mathbf{x}^{\text{IUQ}}$ and $\mathbf{x}^{\text{VAL}}$, respectively. The step 3 of the proposed sequential approach will select the test "furthest away" from the existing tests in $\mathbf{x}^{\text{IUQ}}$. By choosing the test whose input setting is in the unfilled zone of the existing test domain, we aim at extracting the most information for $\boldsymbol{\theta}^{\text{Posterior}}$ using only a relatively small number of $N_{\text{IUQ}}$.

### 3.3. Method to select initial set for validation

In our improved modular Bayesian approach outlined in Figure 1, the computer code output $y^{\text{M}}(\mathbf{x}, \boldsymbol{\theta})$ is first obtained at the input settings of all the tests $\mathbf{x}^{\text{test}}$, with the calibration parameters fixed at nominal values or prior mean values $\boldsymbol{\theta}^0$. The simulation results are denoted as $y^{\text{M}}(\mathbf{x}^{\text{test}}, \boldsymbol{\theta}^0)$. Then $\mathbf{x}^{\text{VAL}}$ and $\{y^{\text{E}}(\mathbf{x}^{\text{VAL}}) - y^{\text{M}}(\mathbf{x}^{\text{VAL}}, \boldsymbol{\theta}^0)\}$ are used as training inputs and output respectively to fit a GP emulator called "GPbias". Evaluating "GPbias" at $\mathbf{x}^{\text{IUQ}}$ results in an estimation of the model discrepancy term $\boldsymbol{\delta}(\mathbf{x}^{\text{IUQ}})$, which will enter the likelihood function during MCMC sampling. Such a treatment of model discrepancy has the following implications for TSA:

1. Because generally a larger training sample size results in a more accurate GP model, $N_{\text{VAL}}$ should be relatively larger than $N_{\text{IUQ}}$. That's why we have picked $\alpha = 25\%$ in Step 4 of the sequential approach for TSA. In this way, about 75% of the measurement data will be used for validation. Furthermore, a relatively smaller number of tests for inverse UQ leads to smaller computational cost of MCMC sampling.

2. It was also demonstrated in [5] that GP emulator is very accurate for interpolation but can cause large errors when used for extrapolation. This fact implies that since "GPbias" is trained with $\mathbf{x}^{VAL}$ and evaluated at $\mathbf{x}^{IUQ}$, the inverse UQ domain $\mathbf{x}^{IUQ}$ should be encompassed by the validation domain $\mathbf{x}^{VAL}$ to avoid extrapolation. Note that this does not mean that $\mathbf{x}^{IUQ}$ should be a subset of $\mathbf{x}^{VAL}$. Each test must be either in $\mathbf{x}^{VAL}$ or in $\mathbf{x}^{IUQ}$. If $\mathbf{x}^{IUQ}$ is a subset of $\mathbf{x}^{VAL}$, $\boldsymbol{\delta}(\mathbf{x}^{IUQ})$ will have zero mean and zero variance, given the fact that GP is an interpolator.

The selection of $\mathbf{x}^{VAL,init}$ can be guided by the above two implications, which states that validation tests $\mathbf{x}^{VAL}$ should have a full coverage of the test domain $\mathbf{x}^{test}$. The term "coverage" refers to the extent to which $\mathbf{x}^{VAL}$ explore the test domain (especially the boundaries) of the selected benchmark data. The estimation of the model discrepancy term $\boldsymbol{\delta}(\mathbf{x}^{IUQ})$ will be limited if the tests with which "GPbias" is trained have insufficient coverage of the test domain. The sequential approach for TSA determines the coverage of the test domain based on convex hull. *Convex hull*, also called *convex envelope*, is the smallest convex domain that envelops all the physical experiments. The coverage $\eta_C$ is calculated using the following equation:

$$\eta_C = \frac{\text{Volume}[\Omega(\mathbf{x}^{VAL})]}{\text{Volume}[\Omega(\mathbf{x}^{test})]} \tag{4}$$

Where $\Omega(\mathbf{x}^{VAL})$ is the convex hull of the domain defined by $\mathbf{x}^{VAL}$, $\Omega(\mathbf{x}^{test})$ is the convex hull defined by $\mathbf{x}^{test}$. Function Volume($\cdot$) calculates the volume of convex hull. The procedure to select $\mathbf{x}^{VAL,init}$ is shown in **Algorithm 1**. This algorithm tries to re-order $\mathbf{x}^{test}$ to $\mathbf{u}$. The major steps are briefly described below:

1) Firstly, the volume of the convex hull $\Omega(\mathbf{x}^{test})$ is calculated;

2) Secondly, choose starting tests for $\mathbf{u}$, because at least $(r + 1)$ tests are required to calculate the volume of a convex hull of dimension $r$. Such starting tests can be those whose input settings contain the minimum or maximum values of each design variable. Denote these starting tests as $\mathbf{u}^{start}$.

3) Thirdly, loop through all the remaining tests in $\mathbf{u}^{rest}$, select the one that maximize the coverage ratio if it is added to the current $\mathbf{u}$. Repeat this process until there is no more test in $\mathbf{u}^{rest}$.

Eventually, step 3 results in a set $\mathbf{u}$ which has the same tests with $\mathbf{x}^{test}$. However, these tests are reordered in $\mathbf{u}$ such that for any number $N_{\mathbf{u},start} \leq n \leq N_{test}$, the first $n$ tests in $\mathbf{u}$ has the largest coverage ratio $\eta_C$ among all the other possible combinations.

---

**Algorithm 1**: Selection of experimental tests for initial validation set $\mathbf{x}^{VAL,init}$

1. Computer the volume of the convex hull $\Omega(\mathbf{x}^{test})$ defined by all the tests $\mathbf{x}^{test}$: Volume$[\Omega(\mathbf{x}^{test})]$.
2. Choose starting tests for $\mathbf{u}$, which is denoted as $\mathbf{u}^{start}$.
$$\mathbf{u} = \mathbf{u}^{start}, \quad N_{\mathbf{u}} = N_{\mathbf{u},start}, \quad \mathbf{u}^{rest} = \mathbf{x}^{test}\backslash\mathbf{u}^{start}, \quad N_{\mathbf{u},rest} = N_{test} - N_{\mathbf{u},start}$$
3. while $N_{\mathbf{u}} < N_{test}$

    for $\mathbf{x}^{(k)} \in \mathbf{u}^{rest}, \ k = 1,2,\ldots,N_{\mathbf{u},rest}$, compute:
    $$\eta_C[\Omega(\mathbf{u} \cup \mathbf{x}^{(k)})] = \frac{\text{Volume}[\Omega(\mathbf{u} \cup \mathbf{x}^{(k)})]}{\text{Volume}[\Omega(\mathbf{x}^{test})]}$$
    end

    select $k^* = \underset{k=1,2,\ldots,N_{\mathbf{u},rest}}{\operatorname{argmax}} \eta_C[\Omega(\mathbf{u} \cup \mathbf{x}^{(k)})]$;

    $$\mathbf{u} = \mathbf{u} \cup \mathbf{x}^{(k^*)}, \quad \mathbf{u}^{rest} = \mathbf{u}^{rest}\backslash\mathbf{x}^{(k^*)}, \quad N_{\mathbf{u}} = N_{\mathbf{u}} + 1, \quad N_{\mathbf{u},rest} = N_{\mathbf{u},rest} - 1$$
    end
4. Eventually, all the tests in $\mathbf{x}^{test}$ are re-ordered in $\mathbf{u}$.

---

The initial validation set $\mathbf{x}^{VAL,init}$ will include the first $N_{VAL,init}$ tests in $\mathbf{u}$ such that the coverage ratio $\eta_C$ becomes 1.0. Note that we do not know the value $N_{VAL,init}$ at the beginning. We can plot the coverage ratio $\eta_C$ of the first $n$ tests in $\mathbf{u}$ as a function of $n$, then choose $N_{VAL,init} = n$ as soon as $\eta_C$ becomes 1.0.

### 3.4. Method to select initial set for inverse UQ

The procedure to select $\mathbf{x}^{IUQ,init}$ is shown in **Algorithm 2** and briefly described below:

1) Firstly, remove $\mathbf{x}^{VAL,init}$ from $\mathbf{x}^{test}$: $\mathbf{x}^{rest} = \mathbf{x}^{test} \backslash \mathbf{x}^{VAL,init}$, and $\mathbf{x}^{IUQ,init}$ will be selected from $\mathbf{x}^{rest}$.
2) Secondly, select a desirable number of tests $N_{IUQ,init}$ for $\mathbf{x}^{IUQ,init}$. For example, $N_{IUQ,init} = \lfloor N_{test} * \beta \rfloor$ where $\beta$ is chosen by the user, e.g. $\beta = 0.05$.
3) Thirdly, starting at each of the tests in the test domain $\{\mathbf{x}^{(i)} \in \mathbf{x}^{rest}, i = 1,2,\ldots,N_{rest}\}$, follow the step 3 in **Algorithm 2** to select the rest ($N_{IUQ,init} - 1$) tests for $\mathbf{x}^{IUQ,init,(i)}$.
4) Eventually we will have $N_{rest}$ different sets $\{\mathbf{x}^{IUQ,init,(i)}, i = 1,2,\ldots,N_{rest}\}$, each set includes $N_{IUQ,init}$ experimental tests and the index (i) means the starting test.

---

**Algorithm 2**: Selection of experimental tests for initial inverse UQ set $\mathbf{x}^{IUQ,init}$

1. Remove $\mathbf{x}^{VAL,init}$ from $\mathbf{x}^{test}$: $\mathbf{x}^{rest} = \mathbf{x}^{test} \backslash \mathbf{x}^{VAL,init}$.
2. Decide the size for $\mathbf{x}^{IUQ,init}$: $N_{IUQ,init} = \lfloor N_{test} * \beta \rfloor$ where $\beta$ is chosen by the user (e.g. 0.05).
3. for $\mathbf{x}^{(i)} \in \mathbf{x}^{rest}$, $i = 1,2,\ldots,N_{rest}$:
       $\mathbf{x}^{IUQ,init,(i)} = \mathbf{x}^{(i)}$;
       $n = 1$;
       while $n < N_{IUQ,init}$
           for $\mathbf{x}^{(k)} \in (\mathbf{x}^{rest} \backslash \mathbf{x}^{IUQ,init,(i)})$, $k = 1,2,\ldots,(N_{rest} - n)$
               compute $D^2[\mathbf{x}^{IUQ,init,(i)} \cup \mathbf{x}^{(k)}]$ or $W^2[\mathbf{x}^{IUQ,init,(i)} \cup \mathbf{x}^{(k)}]$;
           end
           select $k^* = \underset{k}{\mathrm{argmin}}\, D^2[\mathbf{x}^{IUQ,init,(i)} \cup \mathbf{x}^{(k)}]$ or $k^* = \underset{k}{\mathrm{argmin}}\, W^2[\mathbf{x}^{IUQ,init,(i)} \cup \mathbf{x}^{(k)}]$;
           $\mathbf{x}^{IUQ,init,(i)} = \mathbf{x}^{IUQ,init,(i)} \cup \mathbf{x}^{(k^*)}$;
           $n = n + 1$;
       end
   end
4. Step 3 results in $N_{rest}$ different sets $\{\mathbf{x}^{IUQ,init,(i)}, i = 1,2,\ldots,N_{rest}\}$, each set includes $N_{IUQ,init}$ tests.

---

All together there are $N_{rest} \times N_{IUQ,init}$ counts of single test appearances, of which many tests will have multiple appearances. The fact that a certain test appears more frequently means that it is likely to be in an unfilled region in the test domain. We then rank the counts of appearance and select the top $N_{IUQ,init}$ tests to form the initial inverse UQ set $\mathbf{x}^{IUQ,init}$. Note that it is possible to have multiple points located in the same unfilled region that are close to each other. Consequently, they may have similar and high counts of appearances. In that case, only one of them should be selected. Finally, the value of $\beta$ should be much smaller than $\alpha$ and large enough so that $N_{IUQ,init} = \lfloor N_{test} * \beta \rfloor$ has a value of at least 2 or 3. The **Algorithm 2** can efficiently identify those tests that are "far away" from other tests. Such tests have a higher possibility to be selected by the sequential approach in Figure 2. However, putting them in the initial set $\mathbf{x}^{IUQ,init}$ will speed up the TSA process.

### 3.5. TSA results for the BFBT problem

Figure 3 shows the increase in the coverage ratio $\eta_C$ of the first $n$ tests in $\mathbf{u}$ as a function of $n$, $N_{\mathbf{u},start} \leq n \leq N_{test}$. The fact that x-axis starts at 6 means that there are $N_{\mathbf{u},start} = 6$ tests in $\mathbf{u}^{start}$ which include all the lower and upper bounds of the four design variables. We found that a coverage ratio of 1.0 is reached with the first 25 tests in $\mathbf{u}$. Therefore, $\mathbf{x}^{VAL,init}$ takes all these 25 tests and $\mathbf{x}^{IUQ,init}$ will be searched for among all the remaining tests.

In this study, we chose $\beta = 0.05$, which means $\mathbf{x}^{IUQ,init}$ contains $N_{IUQ,init} = \lfloor 78 \times 0.05 \rfloor = 3$ tests. Furthermore, we use a value of 0.25 for $\alpha$ so that $\mathbf{x}^{IUQ}$ consists of $N_{IUQ} = \lfloor 78 \times 0.25 \rfloor = 19$ tests. Figure 4 shows the values obtained from the sequential TSA process for the four design variables after TSA. It is obvious that the tests for inverse UQ are distributed evenly in the test domain. In fact, they have the lowest wrap-around $L^2$ discrepancy among all the possible combinations of the same number of tests. Note that the **Algorithm 2** in Section 3.4 is designed to speed up the TSA process. Therefore, different values of $\beta$ do not have a big influence on the TSA results. However, the influence of different values of $\alpha$ requires further investigation.

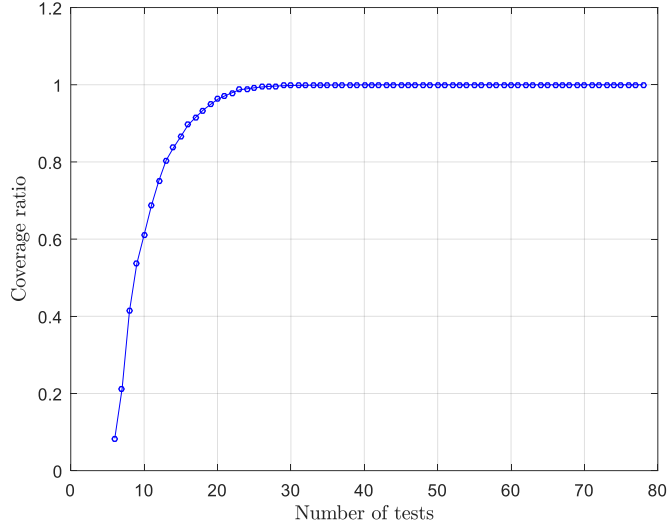

Figure 3: Relation of the coverage ratio with the number of tests in $\mathbf{u}$.

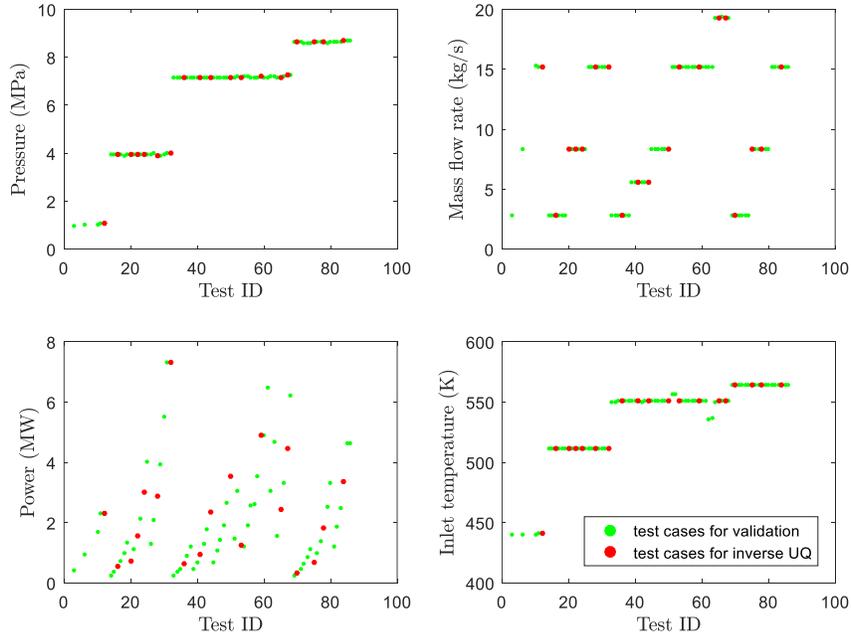

Figure 4: Design variables distribution for $\mathbf{x}^{\text{IUQ}}$ and $\mathbf{x}^{\text{VAL}}$.

## 4. Modeling the TRACE model discrepancy

This section presents the results for the modeling of TRACE model discrepancy $\boldsymbol{\delta}(\mathbf{x})$ (as shown in blocks connected by green arrows in Figure 1). The TRACE code is first evaluated at the input settings of all the tests $\mathbf{x}^{\text{test}}$, with the calibration parameters fixed at nominal values or prior mean values $\boldsymbol{\theta}^0$. The results of model simulations $\boldsymbol{y}^{\text{M}}(\mathbf{x}^{\text{test}}, \boldsymbol{\theta}^0)$ can be compared with $\boldsymbol{y}^{\text{E}}(\mathbf{x}^{\text{test}})$ to obtain TRACE prediction errors. Figure 5 shows the TRACE prediction errors for inverse UQ tests $\{\boldsymbol{y}^{\text{E}}(\mathbf{x}^{\text{IUQ}}) - \boldsymbol{y}^{\text{M}}(\mathbf{x}^{\text{IUQ}}, \boldsymbol{\theta}^0)\}$ and validation tests $\{\boldsymbol{y}^{\text{E}}(\mathbf{x}^{\text{VAL}}) - \boldsymbol{y}^{\text{M}}(\mathbf{x}^{\text{VAL}}, \boldsymbol{\theta}^0)\}$. The x-axis is the test ID ranging from 1 to 86.

Next, design variables $\mathbf{x}^{\text{VAL}}$ and the corresponding TRACE prediction errors $\{\mathbf{y}^{\text{E}}(\mathbf{x}^{\text{VAL}}) - \mathbf{y}^{\text{M}}(\mathbf{x}^{\text{VAL}}, \boldsymbol{\theta}^0)\}$ are used as training inputs and outputs for the GP emulator "GPbias". Maximum Likelihood Estimation (MLE) is used for estimation of the "GPbias" hyperparameters. Evaluating "GPbias" at $\mathbf{x}^{\text{IUQ}}$ results in $\boldsymbol{\delta}(\mathbf{x}^{\text{IUQ}})$ and $\boldsymbol{\Sigma}_{\text{bias}}$, where the former is a mean vector that contains the "expected prediction error" of TRACE at $\mathbf{x}^{\text{IUQ}}$, and the latter is a matrix whose diagonal[3] entries are the variances of the mean vector $\boldsymbol{\delta}(\mathbf{x}^{\text{IUQ}})$. The mean vector $\boldsymbol{\delta}(\mathbf{x}^{\text{IUQ}})$ and covariance matrix $\boldsymbol{\Sigma}_{\text{bias}}$ will enter the following formula during MCMC sampling, which is a part of the likelihood function.

$$p(\mathbf{y}^{\text{E}}, \mathbf{y}^{\text{M}}|\boldsymbol{\theta}) \propto \frac{\exp\left[-\frac{1}{2}[\mathbf{y}^{\text{E}}(\mathbf{x}^{\text{IUQ}}) - \mathbf{y}^{\text{M}} - \boldsymbol{\delta}(\mathbf{x}^{\text{IUQ}})]^{\text{T}}(\boldsymbol{\Sigma}_{\text{exp}} + \boldsymbol{\Sigma}_{\text{bias}} + \boldsymbol{\Sigma}_{\text{code}})^{-1}[\mathbf{y}^{\text{E}}(\mathbf{x}^{\text{IUQ}}) - \mathbf{y}^{\text{M}} - \boldsymbol{\delta}(\mathbf{x}^{\text{IUQ}})]\right]}{\sqrt{|\boldsymbol{\Sigma}_{\text{exp}} + \boldsymbol{\Sigma}_{\text{bias}} + \boldsymbol{\Sigma}_{\text{code}}|}} \quad (5)$$

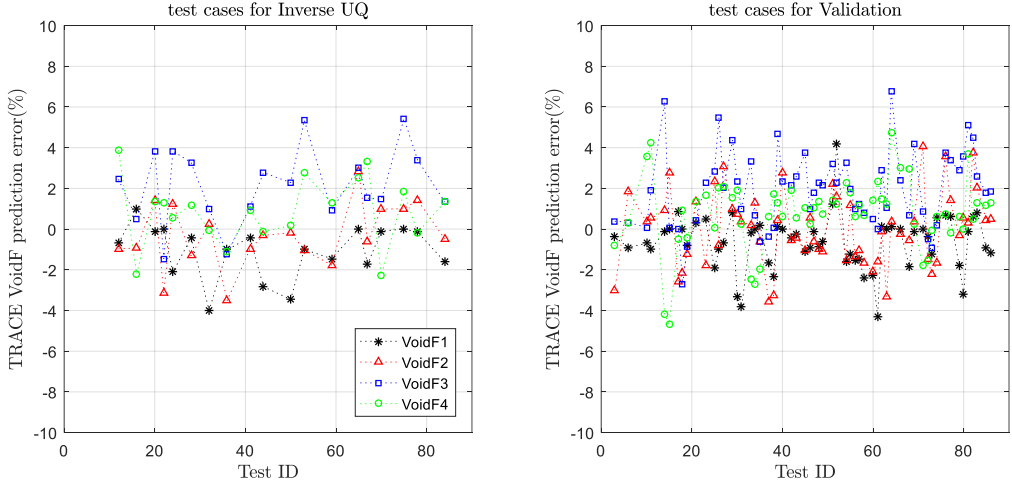

Figure 5: The error in TRACE void fraction simulation for inverse UQ and validation test sets.

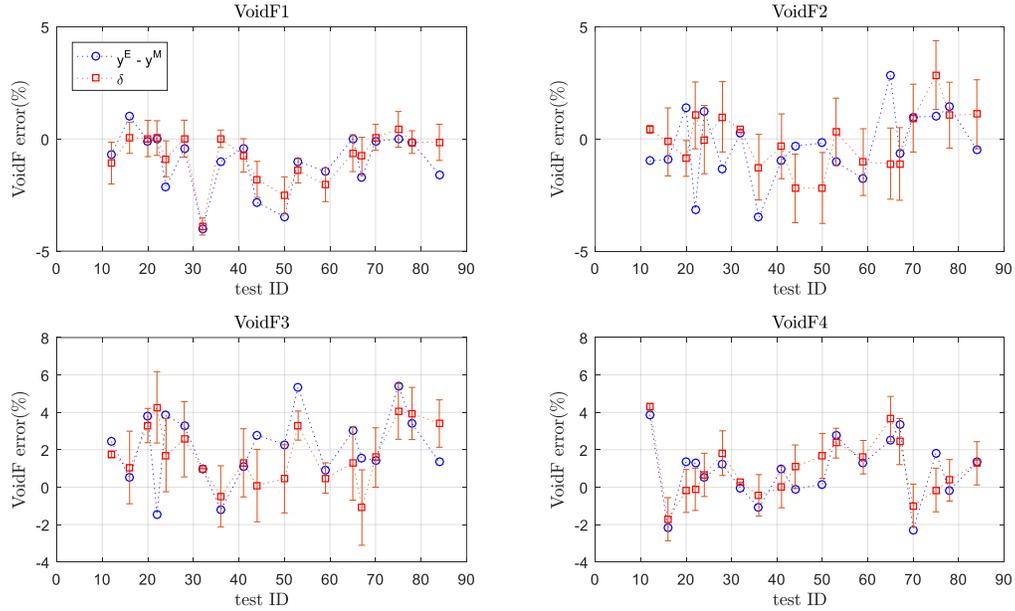

Figure 6: Results of "GPbias" evaluated at $\mathbf{x}^{\text{IUQ}}$.

---

[3] In this work, $\boldsymbol{\Sigma}_{\text{bias}}$ is a diagonal matrix because we do not have enough information for the correlation in different QoIs. But the proposed inverse UQ approach can readily be extended to incorporate such correlations once available.

Figure 6 shows the comparison of the "actual TRACE prediction errors" $\{y^E(x^{IUQ}) - y^M(x^{IUQ}, \theta^0)\}$ and the "expected or interpolated TRACE prediction errors" $\delta(x^{IUQ})$. Also note that the standard deviations $\sqrt{MSE[\delta(x^{IUQ})]}$ are plotted as error bar. Depending on how these two errors compare in these figures, the following different cases can be identified:

1) When $y^E - y^M \approx \delta(x^{IUQ})$, and $\sqrt{MSE[\delta(x^{IUQ})]} \approx 0$, nothing enters Formula (6), so this inverse UQ test is "non-informative". For example, test ID 32 for VoidF3.

2) When $y^E - y^M \approx \delta(x^{IUQ})$, and $\sqrt{MSE[\delta(x^{IUQ})]} \not\approx 0$, $\sqrt{MSE[\delta(x^{IUQ})]}$ enters the denominator of Formula (6), this test is said to be "informative". For example, test ID 70 for VoidF2 and test ID 84 for VoidF4.

3) When $y^E - y^M \not\approx \delta(x^{IUQ})$, this test is "very informative" because substantial information enters the numerator of Formula (6). This is true for most inverse UQ tests.

## 5. Building and Validating GP Metamodel for TRACE

In this section, another GP emulator called "GPcode" is built for TRACE (as shown in blocks connected by purple arrows in Figure 1). Even though TRACE simulation for the BFBT benchmark in the present study is not very expensive (each TRACE simulation takes around 41 seconds), we use it as a placeholder for more computationally prohibitive codes. For those expensive codes there will only be a limited number of runs (a few hundred) available for follow-on analysis. Moreover, 50,000 MCMC samples require the same number of TRACE full model evaluations, which could still take around 24 days with a single processor, making the application of GP emulator necessary.

### 5.1. Building the GP metamodel

The major difference between "GPcode" and "GPbias" is shown in Table 2. "GPbias" uses $x^{VAL}$ as inputs and $\{y^E(x^{VAL}) - y^M(x^{VAL}, \theta^0)\}$ as outputs for training, while "GPcode" uses $(x^{IUQ}, \theta^{Prior})$ as inputs and $y^M(x^{IUQ}, \theta^{Prior})$ as outputs because it is intended to serve as a surrogate model for TRACE. The prior $\theta^{Prior}$ are random parameters uniformly distributed over the ranges shown in Table 1. Such non-informative priors are used to reflect our ignorance about $\theta$. The prior ranges are very important since during MCMC sampling, any trial walk outside the prior ranges will have a zero acceptance probability, meaning that posterior samples will never fall beyond the prior ranges. If the prior ranges are too limited and do not include the "high-probability values" of $\theta$, inverse UQ can never explore these values.

Table 2: Training inputs and outputs for "GPbias" and "GPcode"

|  | GPbias | GPcode |
|---|---|---|
| Training inputs | $x^{VAL}$ | $(x^{IUQ}, \theta^{Prior})$ |
| Training outputs | $y^E(x^{VAL}) - y^M(x^{VAL}, \theta^0)$ | $y^M(x^{IUQ}, \theta^{Prior})$ |

Another notable difference is that, "GPbias" only uses existing values of $x^{VAL}$ and does not perform a design of computer experiments at other $x$ values where no measurement data exist. However, "GPcode" needs to be built with an experimental design of $\theta$ following the distributions of $\theta^{Prior}$, while keeping $x^{IUQ}$ fixed. For every test in $x^{IUQ}$, we generated $N_{design}$ design samples of $\theta$ using maximin Latin Hypercube Sampling (LHS). At this moment we do not know how many design samples are enough to construct an accurate "GPcode" metamodel for TRACE, we tried $N_{design}$ = 3, 6, 9, 12, 15 and 20. Therefore, for every $N_{design}$ value, "GPcode" is trained with $N_{IUQ} \times N_{design}$ samples. Again we use MLE to estimate the "GPcode" hyperparameters.

### 5.2. Validating the GP metamodel

In this work, to evaluate the accuracy of "GPcode" we calculate the predictivity coefficients ($Q_2$) and Leave-One-Out Cross Validation (LOOCV) errors. Table 3 shows the convergence of these two indicators with different $N_{design}$ values. Figure 7 and 8 visualize the results in Table 3. It is apparent that with larger values for $N_{design}$, $Q_2$ values quickly converge to 1.0 for all four QoIs, and LOOCV errors decrease towards zero. In [27] the authors argued that in the literature, a metamodel with $Q_2$ value above 0.7 is often considered as a satisfactory approximation of the full model. In this work, we used a more stringent criterion and require the $Q_2$ values to be above 0.95. Eventually we pick $N_{design} = 20$ to build the "GPcode" emulator. It can be readily used to replace TRACE during MCMC sampling since its accuracy has been proven. The actual cost for training of "GPcode" is $N_{IUQ} \times N_{design} = 19 \times 20 = 380$ TRACE full model runs. At any input setting $\mathbf{x}^*$, "GPcode" metamodel can be evaluated about 23 times per second (0.044 seconds per evaluation). As a comparison, TRACE full model takes about 41 seconds for every evaluation.

Table 3: Predictivity coefficients and LOOCV errors for each design

| Training sample size for each test case | Predictivity coefficient | | | | LOOCV error | | | |
|---|---|---|---|---|---|---|---|---|
| | VoidF1 | VoidF2 | VoidF3 | VoidF4 | VoidF1 | VoidF2 | VoidF3 | VoidF4 |
| 3 | 0.0636 | 0.0456 | 0.0077 | 0.0870 | 2325.10 | 11159.00 | 14174.00 | 2797.40 |
| 6 | 0.6228 | 0.9482 | 0.9017 | 0.9443 | 67.41 | 26.84 | 51.71 | 20.05 |
| 9 | 0.9091 | 0.9716 | 0.9814 | 0.9829 | 12.78 | 15.14 | 9.31 | 6.02 |
| 12 | 0.9008 | 0.9564 | 0.9908 | 0.9978 | 13.07 | 23.89 | 4.59 | 0.77 |
| 15 | 0.9494 | 0.9670 | 0.9953 | 0.9971 | 6.92 | 17.77 | 2.34 | 1.01 |
| 20 | 0.9698 | 0.9841 | 0.9930 | 0.9959 | 4.26 | 8.58 | 3.50 | 1.43 |

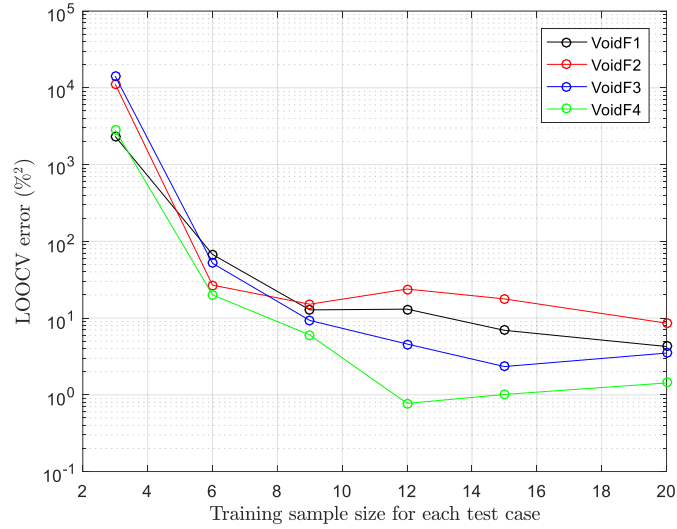

Figure 7: Convergence of LOOCV errors.

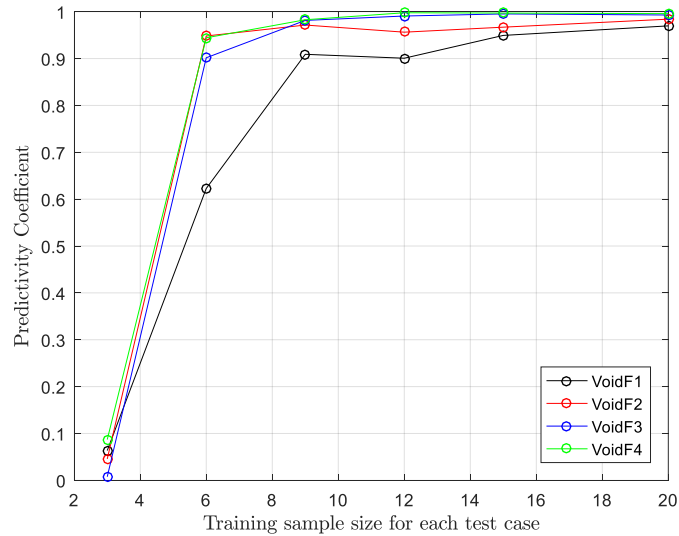

Figure 8: Convergence of predictivity coefficients.

The increase of $Q_2$ values and decrease of LOOCV errors are expected to be monotonic. However, the results in Figure 7 and 8 show that the evolution of these two indicators are not monotonic. This is caused by the randomness of the design of computer experiments. Our conclusion will not be affected as long as the overall trend is consistent with expected monotonic behavior.

## 6. Results for inverse UQ

In this section, we present the results for posteriors by MCMC sampling (as shown in blocks connected by red arrows in Figure 1). The measurement error is assumed to be 2% relative to BFBT void fraction data, which is needed for the variance term $\Sigma_{exp}$ in the posterior. Such a value is reported in BFBT benchmark specification [22]. Note that the benchmark only provides measurement noise related to X-ray CT scanner (VoidF4). We assign the same experimental error for X-ray densitometer (VoidF1, VoidF2 and VoidF3) as there is no better choice.

### 6.1. Posterior samples from MCMC sampling

We used the adaptive MCMC sampling approaches described in [23] which were based on the methods developed in [28]. Fifty thousand samples were generated using "GPcode". It took about 36.7 core-minutes using "GPcode" with a current generation Intel CPU, which would otherwise take about 23.7 core-days using TRACE full model with the same processor. The first 10,000 samples were discarded as burn-in and then every 10$^{th}$ sample were kept from the remainder for thinning of the chain, leaving us with 4,000 samples. Thinning was performed to reduce auto-correlation among the samples. We also generated another MCMC chain without considering the model discrepancy, while keeping everything else the same. In this case, $\delta(x^{IUQ})$ and $\Sigma_{bias}$ need to be removed from Formula (5) and the likelihood function now includes the following part:

$$p(y^E, y^M|\theta) \propto \frac{\exp\left[-\frac{1}{2}[y^E(x^{IUQ}) - y^M]^T(\Sigma_{exp} + \Sigma_{code})^{-1}[y^E(x^{IUQ}) - y^M]\right]}{\sqrt{|\Sigma_{exp} + \Sigma_{code}|}} \quad (6)$$

When inverse UQ is performed without considering the model discrepancy, all the actions indicated by green arrows in Figure 1 will be gone. The posterior $\theta^{Posterior}$ is expected to be over-fitted to $y^E(x^{IUQ})$ and we seek to demonstrate that with the current treatment of model discrepancy in Section 4, such over-fitting can be avoided.

Table 4: Posterior statistical moments

| Parameter | With model discrepancy | | | Without model discrepancy | | |
|---|---|---|---|---|---|---|
| | Mean | STD | Mode | Mean | STD | Mode |
| P1008 | 0.6162 | 0.2113 | 0.4967 | 1.5275 | 0.1923 | 1.3651 |
| P1012 | 1.2358 | 0.0890 | 1.0559 | 1.0844 | 0.0592 | 1.038 |
| P1022 | 1.4110 | 0.1833 | 1.4096 | 0.2452 | 0.1153 | 0.2600 |
| P1028 | 1.3385 | 0.1155 | 1.2044 | 1.4746 | 0.0414 | 1.4300 |
| P1029 | 1.2340 | 0.3453 | 1.0675 | 0.4321 | 0.0833 | 0.2700 |

Table 4 shows the posterior statistical moments including mean values and standard deviations (STD) as well as the modes. The mode of posterior samples for a certain calibration parameters is the value that appears most often, which is essentially the Maximum A Posteriori (MAP) estimate. Figure 9 and 10 show the posterior pair-wise joint densities (off-diagonal sub-figures) and marginal densities (diagonal sub-figures) for the five physical model parameters, with and without considering the model discrepancy respectively. The marginal PDFs are evaluated using Kernel Density Estimation (KDE). The x and y axes of joint densities and x axis of the marginal densities are the prior ranges. As it can be seen, the posterior distributions demonstrate a remarkable reduction in input uncertainties compared to their prior distribution. These density plots are also useful for identifying potential correlations between the parameters, as well as the type of marginal distribution for each parameter. For example, highly linear negative correlation is observed between P1008 and P1012. This indicates that in future forward uncertainty propagation studies, these input parameters should be sampled jointly, not independently, so that their correlation is captured. It can also be noticed that most marginal densities have a shape similar to Gaussian or Gamma distributions.

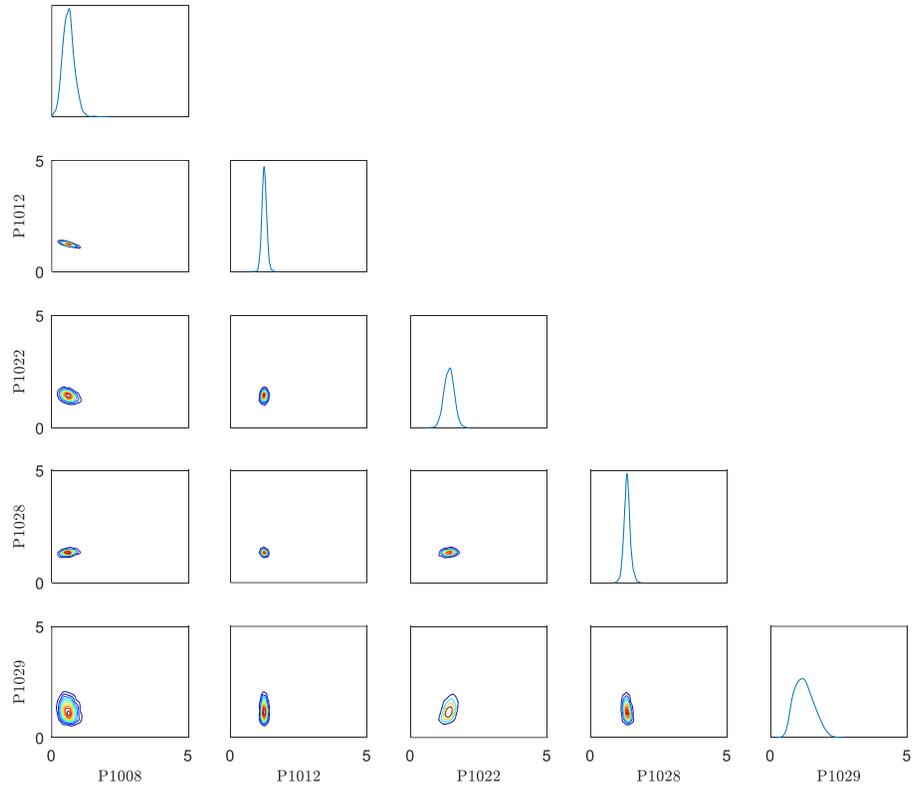

Figure 9: Posterior pair-wise joint and marginal densities when model discrepancy is considered.

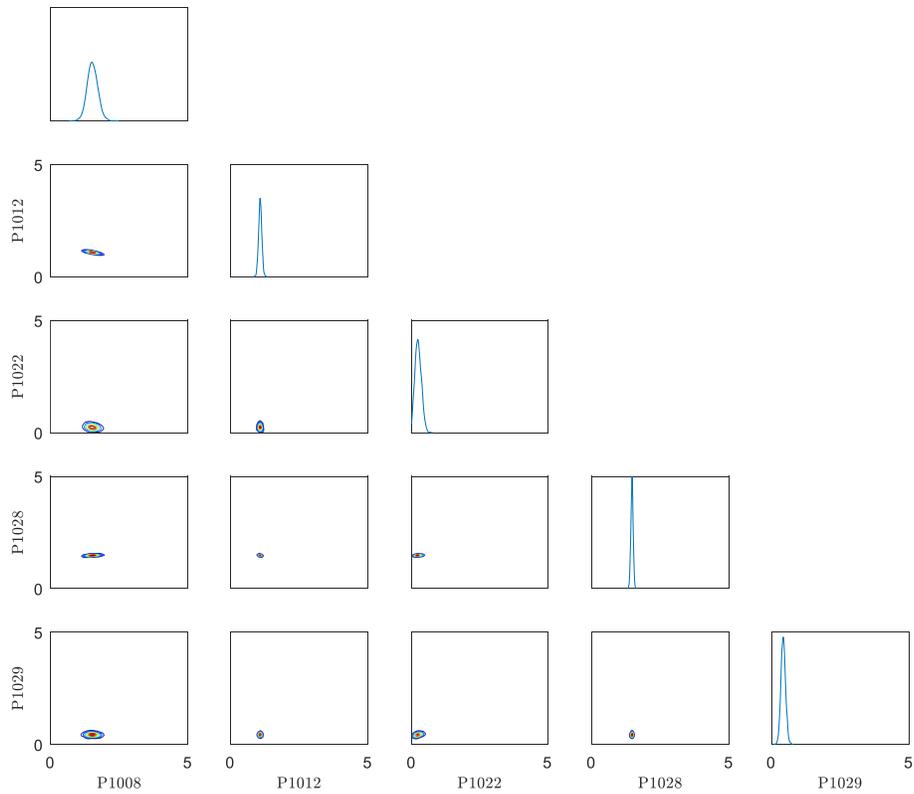

Figure 10: Posterior pair-wise joint and marginal densities when model discrepancy is not considered.

By comparing results with and without model discrepancy in Table 4, and Figures 9 and 10, it can be noticed that when model discrepancy is not considered, the posterior standard deviations are very small and pair-wise joint distributions are more concentrated. This fact is preferable in the sense that more uncertainty reduction is achieved. However, it is also an indication of potential over-fitting. At this point, we do not know which one of the results in Figure 9 and 10 is in fact "closer" to the truth. Therefore, validation (Step 5 in Figure 1) of TRACE based on these posteriors is needed to decide which posterior is more appropriate. To avoid detour in the presentation of the inverse UQ process, results for validation (as shown in by blocks connected by light blue arrows in Figure 1) for the posteriors (with and without mode discrepancy) are omitted in the present paper. A future paper will explain the detailed process of using quantitative validation metrics to validate the achieved posteriors.

**6.2. Fitted posterior distributions**

The validation results (not included in this paper, but can be found in Chapter 7.6 of [29]) show that $\theta^{posterior}$ calculated without considering model discrepancy is over-fitted to the inverse UQ data. Therefore, the posterior $\theta^{posterior}$ achieved when model discrepancy is considered is the preferred results for inverse UQ. To make the posterior distributions more applicable to future forward UQ, sensitivity analysis and validation studies, the posterior samples for each physical model parameter can be fitted to certain well-known distributions, e.g. Gaussian distribution. Figure 11 and Table 5 show the fitted distributions for each physical model parameter and the parameters associated with each distribution, i.e. mean (μ) and standard deviation (σ) for normal distribution, shape ($\alpha$) and scale ($\beta$) parameter for Gamma distribution, non-centrality (s) and scale (σ) parameter for Rician distribution. Gamma and Rician distributions are chosen for certain parameters to keep them strictly positive. All the fitted distributions are accepted by Kolmogorov–Smirnov test at the 5% significance level. Figure 12 shows that good agreement can be achieved between the empirical cumulative distribution function (CDF) and fitted distribution CDF for every physical model parameter. Finally, Table 6 reports the correlation coefficients between the five calibration parameters.

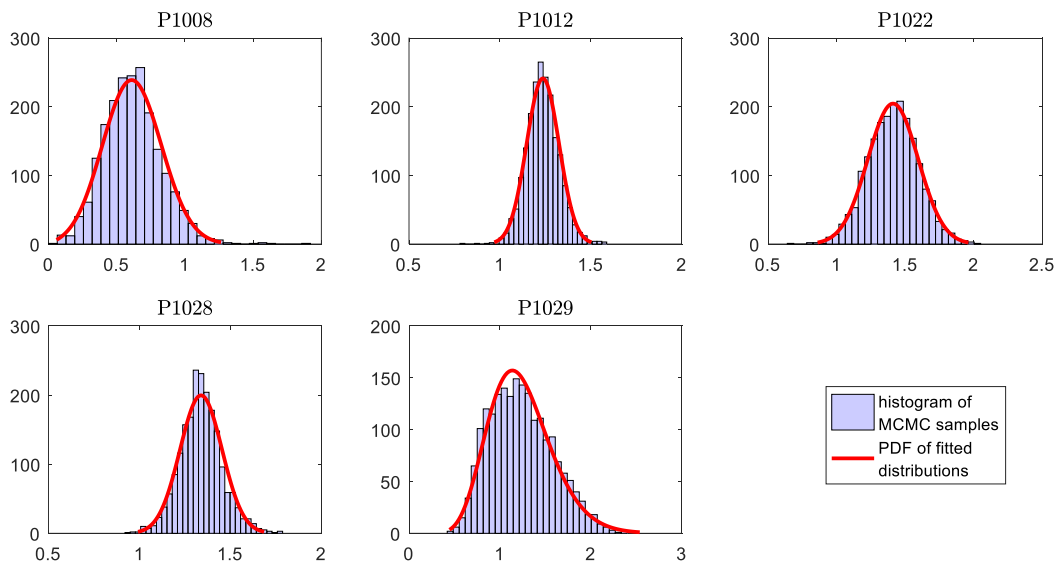

Figure 11: Fitted posterior marginal probability densities

Table 5: Fitted distribution type and distribution parameters

| Parameter | Distribution type | Distribution parameter 1 | Distribution parameter 2 |
|---|---|---|---|
| P1008 | Rician | $s = 0.5709$ | $\sigma = 0.2218$ |
| P1012 | Gaussian | $\mu = 1.2358$ | $\sigma = 0.0890$ |
| P1022 | Gaussian | $\mu = 1.4110$ | $\sigma = 0.1833$ |
| P1028 | Gaussian | $\mu = 1.3385$ | $\sigma = 0.1155$ |
| P1029 | Gamma | $\alpha = 12.6511$ | $\beta = 0.0975$ |

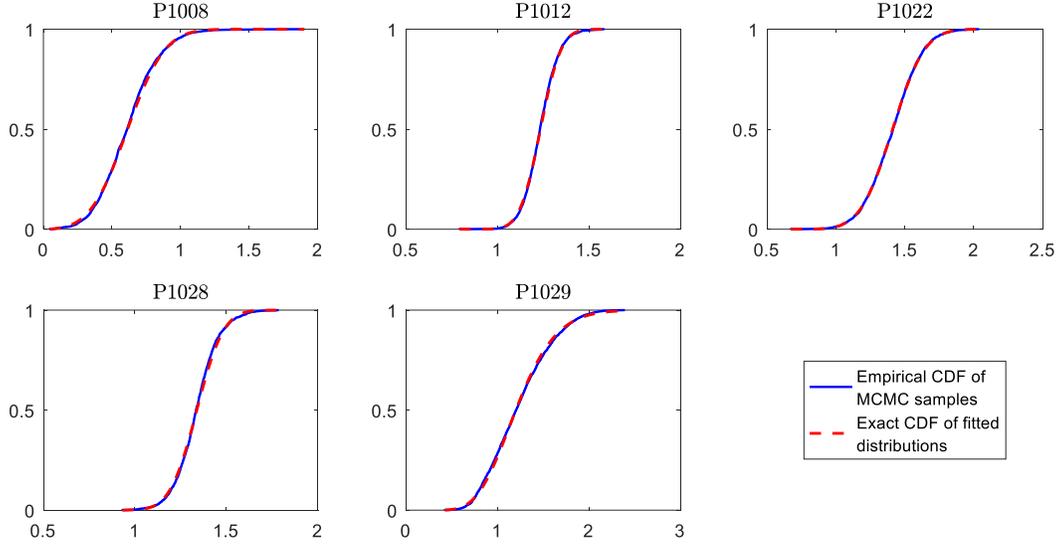

Figure 12: Comparison of empirical CDFs and fitted distribution CDFs.

Table 6: Correlation coefficients of five physical model parameters

|       | P1008   | P1012   | P1022  | P1028   | P1029  |
|-------|---------|---------|--------|---------|--------|
| P1008 | 1.0000  |         |        |         |        |
| P1012 | -0.8338 | 1.0000  |        |         |        |
| P1022 | -0.3543 | 0.0969  | 1.0000 |         |        |
| P1028 | 0.3264  | -0.2121 | 0.1978 | 1.0000  |        |
| P1029 | -0.1941 | 0.0251  | 0.4047 | -0.2246 | 1.0000 |

## 7. Conclusions

In this paper, we applied the improved modular Bayesian approach for inverse UQ developed in a companion paper [5] for TRACE physical model parameters using the BFBT steady-state void fraction data. Inverse UQ aims to quantify the uncertainties in calibration parameters while achieving consistency between code simulations and physical observations. Inverse UQ always captures the uncertainty in its estimates rather than merely determining point estimates of the best-fit input parameters. Also, we developed a sequential approach for efficient allocation of test source (given experimental data) for inverse UQ and validation. This sequential TSA methodology first select tests that has a full coverage of the test domain to avoid extrapolation of model discrepancy term when evaluated at input setting of tests for inverse UQ. Then it selects tests that tend to reside in the unfilled zones of the test domain for inverse UQ, so that inverse UQ can extract the most information for posteriors of calibration parameters using only a relatively small number of tests.

In the Bayesian inference framework, we also considered the model discrepancy term, which in this work was represented by a GP emulator "GPbias". The numerical results demonstrated that such treatment of model discrepancy can avoid over-fitting. Furthermore, we constructed a fast-running and accurate GP emulator "GPcode" to replace TRACE full model during MCMC sampling. The number of TRACE runs is reduced by about two orders of magnitude (from 50,000 to less than 500), and the MCMC sampling time is reduced from about 23.7 core-days to 36.7 core-minutes.

This research addresses the "lack of input uncertainty information" issue about TRACE physical input parameters, which has been often ignored or described using expert opinion or user self-assessment previously. The resulting posteriors of TRACE physical model parameters can be used to replace expert opinions in future uncertainty and sensitivity studies of TRACE code for nuclear reactor system design and safety analysis.

There are some limitations in the current application that need further investigations:

1) Inverse UQ only takes 5 of the 36 TRACE physical model parameters for this study, because only these 5 parameters are significant for BFBT benchmark problem. To inversely quantify the uncertainties in other parameters, different benchmark data are required. For example, parameters P1034 and P1035 will need experimental data that involves reflooding. However, by separated inverse UQ for different groups of parameters, the correlations between the groups will not be quantified. For instance, if parameter groups A and B are significant to benchmarks A and B respectively, then inverse UQ will only result in correlations of parameters within each group but not across the two groups.

2) The current problem has four-dimensional QoIs (VoidF1 - VoidF4) which are supposed to be correlated. However, because we do not have enough information to quantify the correlations between them, they are assumed to be independent to each other when constructing the multi-dimensional GP emulator "GPbias" and "GPcode". As a result, all the three components of the covariance matrix of the likelihood function $\mathbf{\Sigma}_{\text{exp}} + \mathbf{\Sigma}_{\text{bias}} + \mathbf{\Sigma}_{\text{code}}$ are diagonal matrices. The inverse UQ study will be more solid if the correlation structure of the QoIs can be quantified and incorporated in the evaluation of the likelihood.

3) In this study, we used $\alpha = 25\%$ of the experimental data for inverse UQ and the rest for validation. Even though the sequential algorithm for TSA is rigorous and justified, further numerical investigation is needed to see if different $\alpha$ values can cause significant changes in the inverse UQ results.

4) Finally, there is an important issue for inverse UQ that is not addressed in this work: the "identifiability" problem. In our companion paper [5] we have briefly commented on this issue. Inverse UQ problems are usually ill-posed. Many different combinations of the model discrepancy and parameter variability can account for the same amount of error between code simulation and experimental observation, making the true values of the calibration parameters not "identifiable". A future work will present the investigation results on the identifiability issue using the improved modular Bayesian approach.

**Appendix A. List of TRACE Physical Model Parameters**

In this appendix, we provide the full list for 36 physical model parameters implemented in TRACE that can be calibrated. Their IDs, mnemonic names and physical meanings are included in Table A.1. These physical model parameters are referred to as UQ Sensitivity Coefficients in TRACE manual [21]. All of them are Multiplicative factors, with only one exception that parameter $filmTransBoilTMin$ can be Scalar, Additive or Multiplicative. The nominal values for all the Multiplicative factors are 1.0. HTC stands for heat transfer coefficient.

Table A.1. List of 36 physical model parameters implemented in TRACE

| ID | Mnemonic Name | Description |
| --- | --- | --- |
| 1000 | bubSlugLiqIntHTC | Liquid to interface bubbly-slug HTC |
| 1001 | annMistLiqIntHTC | Liquid to interface annular-mist HTC |
| 1002 | transLiqIntHTC | Liquid to interface transition HTC |
| 1003 | stratLiqIntHTC | Liquid to interface stratified HTC |
| 1004 | bubSlugVapIntHTC | Vapor to interface bubbly-slug HTC |
| 1005 | annMistVapIntHTC | Vapor to interface annular-mist HTC |
| 1006 | transVapIntHTC | Vapor to interface transition HTC |
| 1007 | stratVapIntHTC | Vapor to interface stratified HTC |
| 1008 | singlePhaseLiqWallHTC | Single phase liquid to wall HTC |
| 1009 | singlePhaseVapWallHTC | Single phase vapor to wall HTC |
| 1010 | filmTransBoilTMin | Film to transition boiling Tmin criterion temperature |
| 1011 | dispFlowFilmBoilHTC | Dispersed flow film boiling HTC |
| 1012 | subBoilHTC | Subcooled boiling HTC |
| 1013 | nucBoilHTC | Nucleate boiling HTC |
| 1014 | DNB_CHF | Departure from nucleate boiling / critical heat flux |
| 1015 | transBoilHTC | Transition boiling heat transfer coefficient |
| 1016 | gapConductance | Gap conductance coefficient |
| 1017 | fuelThermalCond | Fuel thermal conductivity |
| 1018 | cladMWRX | Cladding metal-water reaction rate coefficient |
| 1019 | fuelRodIntPress | Rod internal pressure coefficient |
| 1020 | burstTemp | Burst temperature coefficient |
| 1021 | burstStrain | Burst strain coefficient |
| 1022 | wallDrag | Wall drag coefficient |
| 1023 | formLoss | Form loss coefficient |
| 1024 | bubblyIntDrag | Interfacial drag (bubbly) coefficient |
| 1027 | dropletIntDrag | Interfacial drag (droplet) coefficient |
| 1028 | bubSlugIntDragBundle | Interfacial drag (bubbly/slug Rod Bundle - Bestion) coefficient |
| 1029 | bubSlugIntDragVessel | Interfacial drag (bubbly/slug Vessel) coefficient |
| 1030 | annMistIntDragVessel | Interfacial drag (annular/mist Vessel) coefficient |
| 1031 | dffbIntDrag | Interfacial drag (dispersed flow film boiling) coefficient |
| 1032 | invSlugIntDrag | Interfacial drag (inverted slug flow) coefficient |
| 1033 | invAnnIntDrag | Interfacial drag (inverted annular flow) coefficient |
| 1034 | tempFlood | Flooding coefficient temperature coefficient |
| 1035 | lengthFlood | Flooding coefficient length coefficient |
| 1036 | invAnnVapWallHTC | Vapor to wall inverted annular HTC |
| 1037 | invAnnLiqWallHTC | Liquid to wall inverted annular HTC |

**Appendix B. Data Correction and Selection for BFBT Benchmark**

The reason for data correction is that the cross-section averaging process of BFBT data was biased. The X-ray densitometers (used to measure VoidF1, VoidF2 and VoidF3) can only capture the void fraction between the rod rows, therefore the measured data only shows the void fraction of a limited area of the subchannel. However, void fraction in the subchannel is not equally distributed as pointed out in [30]. For example, at low void fraction with bubbly flow, the void is concentrated in small bubbles close to the heat surface, while at high void fractions with slug flow, large bubbles are more likely to be located in the subchannel center. Consequently, the void fractions are under predicted at low void fractions and over predicted at high void fractions with the present X-ray densitometers. To resolve this issue, data correction using following correction equation was suggested by [30].

$$\alpha_{\text{corrected}} = \frac{\alpha_{\text{measured}}}{1.167 - 0.001 \cdot \alpha_{\text{measured}}} \quad (7)$$

All the void fractions are in (%) and Equation (7) is recommended for measured void fractions between 20% and 90% (note that VoidF4 is not corrected because it is measured by X-ray computer tomography scanner which does not have the aforementioned limitations of an X-ray densitometer). Figure B.1 shows a comparison of void fraction from BFBT measurements (before and after correction) and TRACE simulations. All 86 test cases are included. Before data correction, the majority of the void fractions are under-predicted especially for VoidF2 and VoidF3. After data correction, the data points are more concentrated close to the diagonal line, indicating good agreements between BFBT and TRACE.

It can be noticed that many of VoidF1 and VoidF2 values are very close to zero. In fact many of them are negative and were not considered in the previous study [23] because negative void fractions are physically wrong. However, in this study we do not abandon these measurements considering that they are only slightly negative. There are also some measurement data that have substantially higher void fractions at lower elevations (e.g. VoidF3 > VoidF4). We decide to remove these non-physical data. Finally, from Figure B.1 (right) we can see some outliers, for instance, the notable VoidF3 by BFBT (~ 20%) for which TRACE simulation is much smaller (~ 4%). These outliers are removed because they have remarkable TRACE simulation errors compared with the majority. They are believed to be caused by measurements failures. Eventually, 8 tests are removed from the original 86 tests. The remaining 78 tests (78*4 = 312 void fraction observations) will be used in the following study.

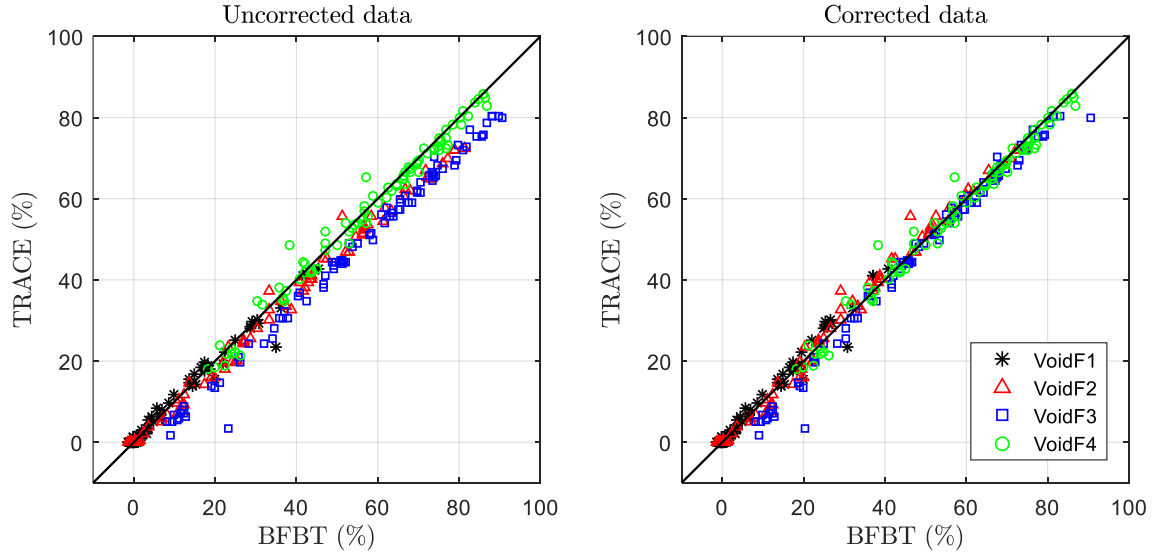

Figure B.1: Comparison of void fractions from BFBT and TRACE for all 86 test cases of assembly 4.